\documentclass[twocolumn,aps,prx,reprint,superscriptaddress,longbibliography]{revtex4-2}

\usepackage{dcolumn}    
\usepackage{bm}         
\usepackage{xfrac}      
\usepackage{physics}    
\usepackage{amsmath}
\usepackage{amssymb}
\usepackage{graphicx}
\usepackage{color}
\usepackage{hyperref}
\usepackage{xcolor}
\usepackage{lipsum}
\usepackage{tabularx}
\usepackage{booktabs}
\usepackage{multirow}
\usepackage{enumitem}
\usepackage{titlesec}
\usepackage{soul}

\hypersetup{
    colorlinks,
    linkcolor={blue!80!black},
    citecolor={blue!80!black},
    urlcolor={blue!80!black}
}

\usepackage[caption=false]{subfig}

\begin{document}

\title{Selective Shuttling of Electrons on Helium Using a CMOS Control Platform}

\author{K.~E.~Castoria}
\affiliation{EeroQ Corporation, Chicago, Illinois, 60651, USA}

\author{H.~Byeon}
\affiliation{EeroQ Corporation, Chicago, Illinois, 60651, USA}

\author{N.~R.~Beysengulov}
\affiliation{EeroQ Corporation, Chicago, Illinois, 60651, USA}

\author{E.~O.~Glen}
\affiliation{EeroQ Corporation, Chicago, Illinois, 60651, USA}


\author{M.~Sammon}
\affiliation{EeroQ Corporation, Chicago, Illinois, 60651, USA}

\author{J.~Pollanen}
\affiliation{EeroQ Corporation, Chicago, Illinois, 60651, USA}

\author{D.~G.~Rees}
\affiliation{EeroQ Corporation, Chicago, Illinois, 60651, USA}

\author{S.~A.~Lyon}
\affiliation{EeroQ Corporation, Chicago, Illinois, 60651, USA}

\date{\today}

\begin{abstract}
Electrons bound to the surface of liquid helium are an emerging quantum computing platform, offering the potential for highly mobile spin qubits that can be manipulated using CMOS-fabricated devices. Here, as a step toward realizing this technology, we demonstrate selective two-dimensional shuttling of electrons across a helium film condensed on the surface of a CMOS control chip. The electrons are moved in packets containing, on average, several tens down to single electrons. We perform CCD-style electron shuttling in any of 128 transport microchannels, each of which links electron storage zones and sensing zones in the 2D plane. Shuttling sequences can be repeated at least 10$^9$ times with no detectable electron loss. The device serves as a prototype quantum information processing platform that is readily scalable to control large monolithically integrated arrays of single electron spins.       
\end{abstract}

\maketitle

\section{Introduction}
\label{sec:intro}

A useful quantum computer will likely require more than $10^5$ physical qubits\cite{Fowler2012ErrorCorrection}. Consequently, qubit platforms compatible with complementary metal-oxide-semiconductor (CMOS) fabrication, such as electron spins in silicon, hold strong promise as scalable quantum technologies\cite{gonzalez-zalbaScalingSiliconbasedQuantum2021}. For such systems, coherent electron shuttling offers an attractive method for coupling distant spin qubits and should allow the efficient execution of quantum algorithms\cite{kunne2024spinbus,sato2025generating}. However, in current solid-state devices the inhomogeneity of the qubit host material and the electron spin-orbit interaction typically make coherent shuttling over even a few $\mu$m along direct paths challenging\cite{jadot2021distant,mills2019shuttling,seidler2022conveyor,struck2024spin,noiri2022shuttling}. 

As an alternative to solid-state spins, electrons bound to the surface of liquid helium (eHe) offer ready integration with CMOS-fabricated host devices as well as predicted long spin coherence times and negligible spin-orbit interaction\cite{lyon2006spin}. Attracted to the liquid dielectric by a small image charge but prevented from penetrating by a $\sim 1$~eV potential barrier at the surface, electrons remain trapped $\sim 10$~nm above the helium\cite{GRIMES1978}. Because the helium surface is free of impurities or charge traps the eHe mobility in the parallel plane is extremely high\cite{sommer1971mobility,Iye1980_Mobility} and the electron motion can be precisely controlled using gate electrodes patterned beneath the liquid surface\cite{glasson2001observation,rees2011point}. Indeed, the ultra-efficient shuttling of electron packets on helium in around 100 parallel transport channels has already been demonstrated using CMOS foundry-fabricated devices  \cite{Bradbury2011Clocking,takita2012spatial,Takita2014Isolating}. Here we extend this work, leveraging the sophistication of the CMOS architecture and the optimization of our electron control scheme to selectively address electron packets at specific locations in a 2D charge-packet array. This allows us to demonstrate choreographed electron shuttling in any of 128 transport microchannels. By systematically reducing the packet charge density we find evidence that single electrons can be isolated and transported, a key requirement for future quantum processors. Indeed, the trapping and shuttling of electrons in a vacuum demonstrated here is reminiscent of qubit transport in other quantum computing platforms such as trapped ion \cite{pino2021demonstration,moses2023race,delaney2024scalable}and neutral atom systems\cite{barredo2016atom,bluvstein2022quantum,chiu2025continuous} but is performed without the need for complex dynamical electric or optical fields. 

The highly mobile nature of eHe spin qubits should allow quantum gate operations and qubit readout to be performed with all-to-all connectivity by shuttling individual qubits to specific operation zones in a quantum processor. Much of the device architecture and functionality required to achieve this aim is exhibited here with our CMOS control chip. Our measurements are facilitated by a charge-sensing scheme that employs two independent HEMT amplifiers mounted at low temperature in close proximity to the chip\cite{feldman2025sensing}. Each are used for electron packet sensing from two separate groups of sensors on the device. While more sensitive devices such as single electron transistors or superconducting microwave resonators have been used to detect single electrons on helium\cite{Papageorgiou2004,Yang2016,koolstra2019coupling,castoria2025sensing}, HEMT amplifiers offer a versatile alternative with no strict device fabrication requirements, wide measurement bandwidth, off-chip mounting and low cost\cite{blumoff2022fast, mills2022high}. Demonstration of a HEMT amplifier as an effective charge sensor for multiple, individually addressed eHe ensembles is therefore a key result of this work. The efficient and controlled eHe shuttling demonstrated here along with a sensitive and versatile charge readout scheme allows our device to serve as a prototype eHe quantum information processor. 

The paper is arranged as follows: in Section~\ref{sec:Device} we describe the device. In Section~\ref{sec:packet} we describe how electron packets are isolated on the helium surface using underlying gate electrodes. In Section~\ref{sec:sensor} we fully characterize our electron sensing scheme with the aid of finite element modeling (FEM) of the device architecture. In Section~\ref{sec:single} we demonstrate control of the number of electrons per packet and present evidence for single electron control. In Section~\ref{sec:clocking} we describe the biasing scheme used to address specific electron packets on the chip and use it to perform selective electron shuttling. We then summarize our results in Section~\ref{sec:conclusions}.

\section{Wonder Lake Device}
\label{sec:Device}

\begin{figure} 
\centering
	\includegraphics[width=0.9\linewidth]{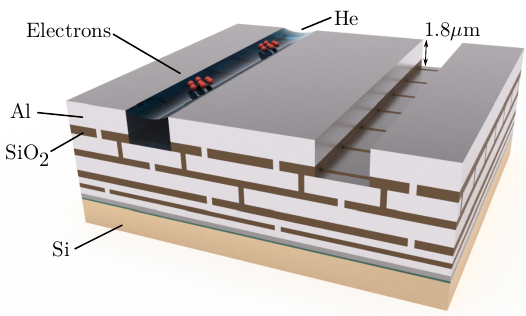}
	\caption{Schematic of the Wonder Lake Device showing surface microchannels and cross section. The surface-state electrons and superfluid helium occupy the channels defined in the uppermost layers of the aluminum interconnects. The channel walls are comprised of 505 nm of silicon oxide and 1.26 $\mu$m of aluminum. The CCD and control electrodes (visible on the right) are patterned on the second aluminum layer from the top. The lower interconnect layers are used for signal routing and shielding.}
\label{fig:Stack}
\end{figure}

\begin{figure*} 
\centering
	\includegraphics[width=0.9\linewidth]{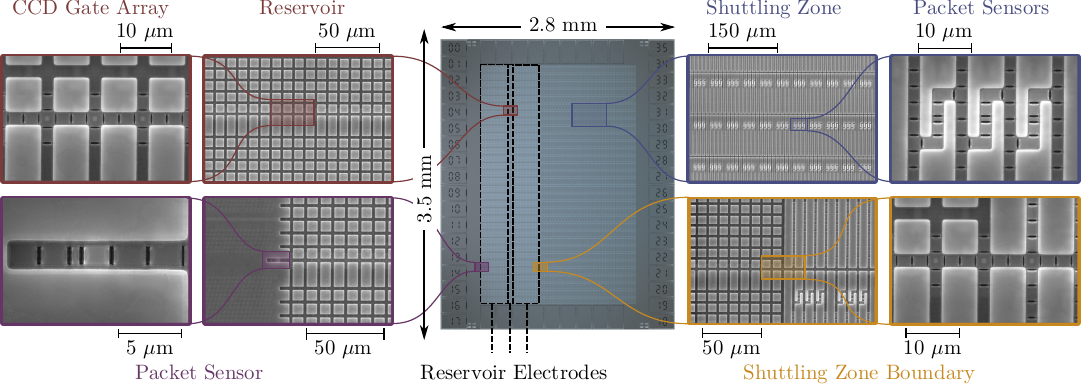}
	\caption{Top-down view of the Wonder Lake device. In the center is an overview of the entire device area. the three large area reservoir electrodes are outlined in black dashed lines. In the four corners are SEM images of (clockwise, from top left) the Reservoir region and its CCD channels, the electron Shuttling Zone and its embedded packet sensors, the transition between the Reservoir region and electron Shuttling Zone, and the electron packet sensors to the left of the Reservoir.}
\label{fig:DeviceOverview}
\end{figure*}

The CMOS control chip, named `Wonder Lake', was fabricated using the SKY130 130 nm CMOS process at the SkyWater Technologies silicon foundry\cite{skywater_sky130_pdk}. The device stack is shown schematically in Fig. \ref{fig:Stack}. A silicon wafer substrate supports five metal layers, separated from one another by insulating layers, with vertical metallic vias used to make connections between them. In a standard CMOS device, these metal layers provide interconnects and power rails for transistors formed in the silicon substrate. Here the upper metal layers are used to create a system of microchannels, which are filled with liquid helium, and gate electrodes that control and detect electrons on the helium surface.

Following foundry fabrication, the only post-processing step required to ready the device for eHe measurements is a fluorine reactive ion etch that removes the sections of the upper insulator layer left exposed after the CMOS fabrication process. This creates a pattern of microchannels on the surface of the chip, defined by the remaining metal electrodes, as shown in Fig. \ref{fig:Stack}. The chip is wire-bonded to a host PCB, which is then hermetically sealed in a copper enclosure and mounted on the cold stage of a 1.2 K pumped helium cryostat. A thin stainless steel tube connects the enclosure to a gas handing system at room temperature which supplies helium to the experiment. When liquid helium is introduced to the device enclosure, a thin ($\sim 30$ nm) superfluid film covers the entire surface of the device. The microchannels then fill with helium due to the capillary action of the liquid\cite{marty1986stability}. A small tungsten filament positioned above the device is used to generate electrons within the enclosure, some of which are trapped on the surface of the helium in the microchannels. The number of electrons collected is controlled by positive voltages applied to the electrodes at the base of the microchannels.

The microchannels are 2.5 $\mu$m wide and, since they are patterned in the uppermost metal layer of the device stack, 1.8 $\mu$m deep. Microchannels of similar dimensions, fabricated in academic clean room facilities, have been widely used to investigate phenomena such as single-eHe transport and electron correlation effects in confined geometries\cite{glasson2001observation,ikegami2012evidence,rees2011point,rees2016structural,kleinbaum2018thermopower}. However, in such fabrication facilities device size and yield can be limited by lithography resolution and particulate contamination. Here, the highly controlled CMOS fabrication process enables the creation of high-resolution microchannel patterns over a large device area, with no detectable particulate contamination or process-induced defects. Additionally, the multilayer device stack allows fabrication of device architectures with a level of complexity not feasible in a standard academic clean-room process.

An overview of the device is shown in Fig.~\ref{fig:DeviceOverview}. The total chip size is 2.8$\times$3.5 mm, with 54 wire bond pads arranged around the perimeter. Of these, 34 were used to perform the eHe measurements described here. The microchannel pattern covers a 1.9$\times$2.9 mm area at the center of the chip. As shown in the expanded images, the left side of the device contains a large grid-like array of interconnected vertical and horizontal microchannels. This region is used to store a large number of surface electrons and is therefore referred to as the $Reservoir$. The square islands formed in the upper metal layer, each 6.5$\times$6.5 $\mu$m, are held at -6 V, while a positive dc voltage (typically +5 V) is applied to the metal layer at the bottom of the microchannels. To monitor the presence of electrons in the reservoir, the lower metal layer is split into three electrodes, indicated with black dashed lines in the center of Fig.~\ref{fig:DeviceOverview}. Applying a 1 V$_{\mathrm{rms}}$ ac driving voltage to one of the outer electrodes induces electron motion within the Reservoir microchannel network. This in turn induces an ac current in the other outer electrode, which is measured using a lock-in amplifier~\cite{sommer1971mobility}. This ac current signal is used to verify charging of the helium surface after the filament discharge, and to periodically check that the electron density in the Reservoir is maintained during the electron shuttling measurements described below. By sweeping the voltage on the central electrode and measuring the voltage required to suppress this transport signal, we determine the electrochemical potential in the Reservoir region, $\phi_\mathrm{e}$, to be around 4.92~eV~\cite{GLASSON2000}.  We calculate that given the surface area of the microchannel network and this electrochemical potential, approximately $3\times10^6$ electrons are stored in the Reservoir in total. 

The Reservoir is intersected by 34 horizontal channels with square gate electrodes patterned along the bottom. Every third gate electrode is connected together through the underlying metal layers. This wiring scheme allows us to use a 3-phase voltage sequence to shuttle charge `packets', typically containing several tens of electrons, laterally along the channels in a manner similar to charge clocking in CCD-based image detectors\cite{janesick2001scientific}. The loading of these packets from the Reservoir into these  arrays is described in more detail in Section~\ref{sec:packet}. A typical 3-phase shuttling sequence uses voltages of +5 V and –4 V, with the upper electrode held at –6 V. We are able to perform electron shuttling at frequencies up to hundreds of kHz, as limited by the high-frequency filtering of our cryostat wiring. The shuttling can be performed for many hours (during which in excess of $10^9$ clocking operations are completed) with no observable loss of charge from the electron packets, during which time the electrons typically travel several tens of km. 

The left-hand edge of each CCD gate array is extended to provide sites for the capacitive sensing of electron packets. One of these packet sensors is shown in the lower-left corner of Fig.~\ref{fig:DeviceOverview}. Each packet sensor comprises several gate electrodes used for electron control and sensing. These electrodes are identically arranged and connected in parallel across all 34 packet sensors. As a result, electron control operations are performed simultaneously in all 34 packet sensors and the charge readout signal is the sum of the signals from each, providing a boost in signal-to-noise ratio. Operation of the packet sensor is described in detail in Section~\ref{sec:sensor}, and its use in determining the number of electrons per charge packet is described in Section~\ref{sec:single}. 

The horizontal CCD gate arrays extend from the packet sensors, across the Reservoir, to the electron \emph{Shuttling Zone} situated on the right side of the device. The Shuttling Zone features a microchannel unit cell, which is repeated 32 times in the horizontal direction and 34 times in the vertical direction. A set of the unit cells is shown in the upper-right corner of Fig.~\ref{fig:DeviceOverview}. Each unit cell comprises four interlinked microchannels that feature gates for storing, addressing, and sensing electron packets. The boundary between the Reservoir region and the Shuttling Zone is shown in the lower-right corner of Fig.~\ref{fig:DeviceOverview}. As described in Section~\ref{sec:clocking}, electrons can be shuttled deterministically within the Shuttling Zone geometry, both inside individual unit cells and between locations in different unit cells. This allows us to perform transport operations such as moving electrons from storage zones to packet sensors, or bringing electrons together at specific locations for set periods of time to choreograph future two-qubit gate operations. Thus we are able to demonstrate several of the key functionalities ultimately required for using eHe as a mobile qubit platform.

\section{Electron Packet Initialization}
\label{sec:packet}

\begin{figure} 
\centering
	\includegraphics[width=0.9\linewidth]{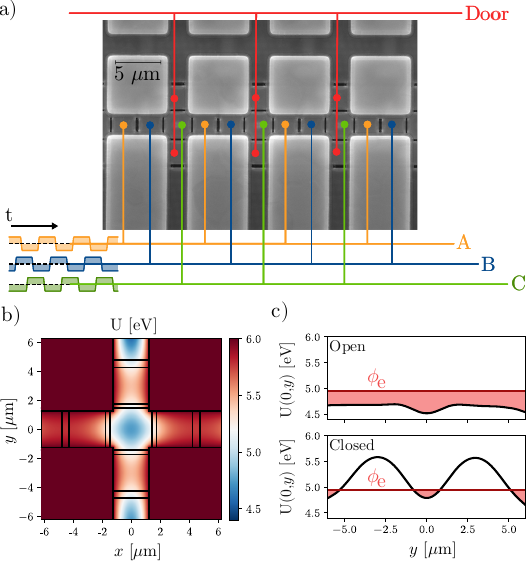}
	\caption{Electron packet initialization. (a) SEM image of the CCD array crossing the Reservoir. Wiring of the three phases to the CCD gates and the Door electrodes is shown via the orange, blue, green, and red wires respectively. The 3-phase voltage pulse sequence applied to the CCD gate lines to move electron packets in the positive $x$ direction is shown on the left as a function of time $t$. (b) The simulated potential energy surface during packet initialization. Here the electrodes are biased such that there are isolated minima atop each of the C phase gates on the CCD array. The Door gates for this simulation are in the ``closed" state (-5V) such that the minima are separated from the reservoir. (c) Line cuts along $x=0$ of the potential energy profile in (b). The upper (lower) panel shows the potential as a black line when the Door gate is in the open (closed) state biased to $+2V$ ($-5V$). The measured electrochemical potential of electrons in the Reservoir $\phi_\mathrm{e}$ is shown with a red line.}
\label{fig:PacketInit}
\end{figure}

In this Section we describe how electron packets are isolated from the extended system of electrons that populates the Reservoir and loaded into the CCD channels. A sequence of gate voltages admits electrons into the CCD channel from the Reservoir before creating a potential barrier between them. As a result, packets containing small numbers of electrons remain isolated along each CCD channel. These packets can then be shuttled left or right along the CCD channel by applying phase shifted voltage sequences, as depicted in the lower left corner of Fig.~\ref{fig:PacketInit}(a). During this sequence, the three phases alternate between +2~V and -5~V at 300~kHz.

An image of the intersection between the Reservoir and one of the CCD channels is shown in Fig.~\ref{fig:PacketInit}(a). The three voltage phases of the CCD gates are labeled $A$, $B$, and $C$. The C-phase of the CCD gate array (CCD-C) is separated from the Reservoir by gates both above and below which are referred to as the \emph{Reservoir Doors}. Fig.~\ref{fig:PacketInit}(b) depicts the electrostatic potential energy, $U(x,y)$, in the plane of the helium surface, with these Door electrodes closed, as calculated by FEM\cite{zhk}. Here the Reservoir electrode and CCD-C are biased at 2~V, the remaining CCD phases and Reservoir Door electrodes are biased at -5~V, and the top metal is biased at -6~V. For this bias configuration, isolated minima are visible above the CCD-C gates. To fill these areas with electrons, the Reservoir Door electrode is swept to +2~V. Now, the electrochemical potential of the electron system in the Reservoirs is sufficiently high such that the electrons are able to to spill over the Reservoir Doors and fill the region above the CCD-C gates ($\phi_\mathrm{e} > \mathrm{max}[U(0,y)]$). Sweeping the Door electrodes back to -5~V, some electrons remain isolated above the CCD-C gates creating packets to be clocked along the channel using the 3-phase CCD voltage sequence. Line cuts of the potential along $x=0$ are seen in Fig.~\ref{fig:PacketInit}(c) for both the open and closed Reservoir Door configurations. 

\section{Electron Packet Sensor Characterization}
\label{sec:sensor}

\begin{figure*} 
\centering
	\includegraphics[width=0.9\linewidth]{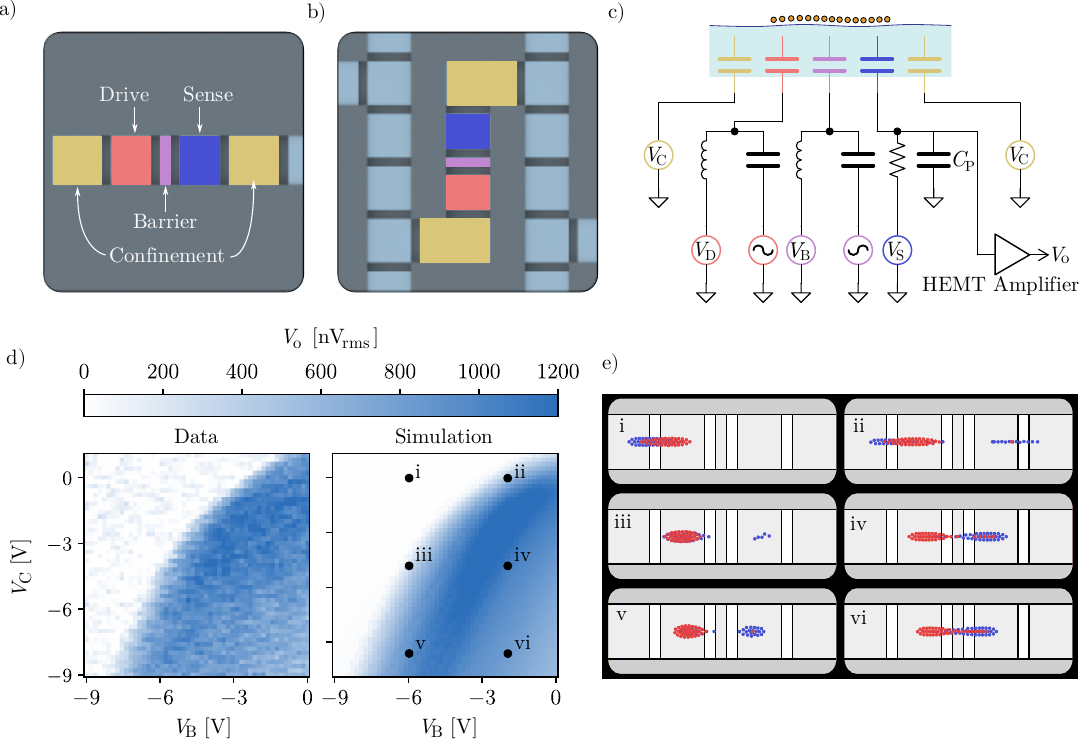}
	\caption{Electron sensor characterization. (a) Schematic of 1 of the 34 packet sensors which populate the left edge of the Reservoir. (b) Schematic of 1 of the 3264 packet sensors which populate the Shuttling Zone. In both cases, the packet sensor consists of a Drive gate, Barrier gate, Sense gate, and two Confinement gates, as described in the main text. (c) Schematic of the charge measurement scheme. Bias tees on the Drive and Barrier electrode allow for the driving and cancellation signals to be superimposed on the applied dc potential. The HEMT amplification circuitry wired to the sense gate amplifies the ac voltage established across the parasitic capacitance to ground, $C_\mathrm{P}$. (d) Experimental and simulated ac sensor voltage $V_\mathrm{O}$, for varying Barrier and Confinement electrode bias and 37 electrons loaded into the sensor. (e) Simulation results showing electron positions for the 37-electron packet for six different dc voltage configurations. The voltage configurations are indicated by the dots in the right panel of (d). The red and blue dots correspond to electron positions with the drive at its most positive (red) and negative (blue) values. The amplitude of the output signal, $V_\mathrm{O}$, is proportional to the number of electrons moving on and off the the sense electrode per cycle.}
\label{fig:SensorExplanation}
\end{figure*}

In this Section we describe how the packet sensors are used to sense small numbers of electrons. As well as the 34 packet sensors on the left edge of the Reservoir (shown schematically in Fig.~\ref{fig:SensorExplanation}(a)), each unit cell in the electron Shuttling Zone also contains 3 identical packet sensors (Fig.~\ref{fig:SensorExplanation}(b)). In this Section we will focus on measurements performed with the 34 packet sensors adjacent to the Reservoir. However, as shown in Section~\ref{sec:clocking}, electron sensing in the Shuttling Zone can be performed in an identical manner.

A schematic image of one packet sensor is shown in Fig.~\ref{fig:SensorExplanation}(a). The sensor consists of five independent gate electrodes patterned at the bottom of the microchannel. At the center, a thin electrode referred to as the \emph{Barrier} is used to control electron transport along the packet sensor. To the left and right of the Barrier are the \emph{Drive} and \emph{Sense} electrodes, respectively, which are used to perform the charge measurement. At the outermost left and right sides of the packet sensor are \emph{Confinement} electrodes, which are used to modulate the effective area, and thus the electrochemical potential, of the electron packet. 

After initialization in the CCD array, as described in Section~\ref{sec:packet}, electron packets are clocked one period left using the three-phase voltage sequence. With all sensor electrodes made positive, the left-most electron packet is thereby loaded into the packet sensor. The Confinement electrodes are then set to -9 V to confine the packet above the Drive, Barrier and Sense electrodes.

The charge detection method is outlined schematically in Fig.~\ref{fig:SensorExplanation}(c). A sinusoidal excitation voltage is superimposed on the dc bias applied to the Drive electrode. 
Electrons within the packet sensor move along the helium surface in response to this drive, inducing an image current in the Sense electrode\cite{GRIMES1978,Bradbury2011,takita2012spatial}. We use a cascode HEMT amplifier circuit identical to that described in Ref.~\cite{feldman2025sensing} mounted close to the device at low temperature, together with a lock-in amplifier at room temperature, to measure the voltage $V_{\mathrm{O}}$ established as this ac current flows to ground through a small (19 pF) stray capacitance. Our signal-to-noise ratio is improved by superimposing a second, smaller input signal to one of the other sensor electrodes, as shown in Fig.~\ref{fig:SensorExplanation}(c). This second signal is at the same frequency as the driving signal but nearly opposite in phase. Prior to electron packet loading, the phase and magnitude of this cancellation signal are tuned such that the output signal of the empty packet sensor due to the parasitic crosstalk between the Drive and Sense gates is minimized.
At the experimental temperature, and for the frequencies used in these measurements ($\sim 1$~MHz), our HEMT amplifier circuits show typical noise amplitudes of $\sim10$~nV$/\sqrt{\mathrm{Hz}}$, close to values recorded elsewhere\cite{feldman2025sensing}.  
  
Results of a charge sensing measurement are shown in Fig.~\ref{fig:SensorExplanation}(d). Here, all 34 packet sensors are loaded with electron packets from the Reservoir. Because the clocking sequence used to load each sensor is identical, and the electrochemical potential is the same for all electrons in the Reservoir, we assume the same number of electrons populate each sensor. A 2 V$_\mathrm{pp}$ excitation, of frequency 8.72 MHz, is applied to the Drive electrode. The Sense and Drive electrodes are dc-biased at 0 V and the upper electrode and CCD electrodes held at -6 V and -9 V, respectively. The bias voltages on the Barrier and Confinement electrodes are then swept from -9 V to -1 V and -10 V to 0 V, respectively. The values of $V_{\mathrm{O}}$ recorded during these gate voltage sweeps are shown in the left panel of Fig.~\ref{fig:SensorExplanation}(d). The plot reveals two distinct regions: for positive Barrier and negative Confinement voltages, electrons move above the Sense electrode in response to the Drive voltage resulting in a large $V_{\mathrm{O}}$, while for the opposite case $V_{\mathrm{O}}$ is small indicating a suppression of electron motion.

To understand this result we use the FEM analysis of the electrostatic potential within the packet sensor to perform numerical simulations of electron transport.
We begin by assuming that a charge packet containing $N_\mathrm{e}$ electrons is initialized in the packet sensor. For each DC voltage configuration, FEM is used to calculate the electrostatic potential within the sensor, and the equilibrium electron positions that minimize the total potential energy are determined numerically~\cite{Yang2016,zhk}. The procedure is then repeated for the maximum and minimum voltages applied to the Drive electrode during each ac voltage cycle ($V_\mathrm{Drive} = +V_\mathrm{pp}/2$ and $-V_\mathrm{pp}/2$). Between each step, the electron system is allowed to relax to a new equilibrium configuration, simulating the response of the electron packet 
to the ac drive. For each electron configuration the charge induced on the Sense electrode by the electron ensemble is determined using the dimensionless coupling constant $\alpha(x,y)$, which relates the voltage induced across the helium surface for a 1~V potential applied to the Sense electrode. Having obtained the coordinates for each electron $i$ in the sensor, the total induced charge on the Sense electrode is given by $q_\mathrm{s} = e\sum\limits_{i} \alpha(x_i,y_i)$, where $e$ is the elementary charge. The variation of the induced charge between the ac voltage extremes, $\Delta q_\mathrm{s}$, allows estimation of the root-mean-square packet sensor voltage signal as    

\begin{equation} \label{eq:amplifier_amplidue}
V_\mathrm{O} = \frac{1}{2\sqrt{2}}\frac{G_\mathrm{\omega} N_\mathrm{S}}{2\pi C_\mathrm{P}} \Delta q_\mathrm{s} ~,
\end{equation}

\noindent where $G_\mathrm{\omega} = 6$ is the amplifier gain at the measurement frequency, $N_\mathrm{S} = 34$ is the number of sensors being measured in parallel and $C_\mathrm{P} = 19$ pF is the capacitance of the Sense electrode to ground. We repeat this simulation process for different $N_\mathrm{e}$ to determine the packet size that best fits the experimental data. 

As an example, the results of this simulation for $N_\mathrm{e}=37$ are shown in the right panel of Fig.~\ref{fig:SensorExplanation}(d). For this electron number we find excellent agreement with the experimental data. In Fig.~\ref{fig:SensorExplanation}(e) we show the simulated electron positions for six different dc voltage configurations, which reveal details of the electron distribution and transport. Positions at the positive and negative extremes of the voltage drive are shown with red and blue dots respectively. For the most negative Barrier and most positive Confinement voltages (region i), electrons cannot overcome the potential barrier at the center of the packet sensor and remain localized on one side of the Barrier. As the Barrier is made more positive, and the Confinement more negative, electrons begin to spill over the barrier, causing a sharp increase in $V_\mathrm{O}$ (regions ii and iii), before eventually becoming strongly confined above the Barrier electrode (region vi). A maximum in the simulated ac signal occurs when the confinement is not too strong and the lateral electron motion within the packet sensor is maximal, a feature also observed in the experiment (regions iv and v). The excellent agreement between the experimental and simulated ac signal characteristics, and the boundary marking the onset of electron motion, confirms an accurate understanding of the electron motion within the packet sensor and of the number of electrons populating each charge packet. 

We note that for extended eHe systems moving above a system of electrodes, screening of the electron sheet can lead to significant phase shifts and damping of voltage excitations. In such cases, eHe transport should be analyzed by treating the electron system as a transmission line with distributed resistance, capacitance and inductance\cite{mehrotra1987analysis}. Here, despite the relatively high driving frequency, we do not consider such effects. Evaluation of the 2D skin depth $\delta_\mathrm{0} = (2n_\mathrm{s}e\mu /\omega C') ^{1/2}$ \cite{lea1991ac} yields $\delta_\mathrm{0} \approx 1$~mm, a distance much larger than our electron packet size (here $n_\mathrm{s} \approx 1\times10^{13}$~m$^{-2}$ is the typical electron surface density, $\mu \approx 10$~m$^2$/V$\cdot$s is the electron mobility at 1.3 K\cite{Iye1980_Mobility}, $\omega$ is the driving frequency in rads/s and $C' \approx 5$~$\mu$F/m$^{2}$ is the capacitance per unit area derived from a simple parallel-plate capacitor model of an electron system a distance $d$ from the underlying electrodes). This confirms that the electron packets can respond to the driving voltage much faster than than the ac period, and justifies our numerical treatment of the electron system in the dc limit where the packet distribution is in electrostatic equilibrium with the applied gate potentials. 

\section{Evidence for Single-Electron Packet Initialization}
\label{sec:single}

\begin{figure*} 
\centering
\includegraphics[width=0.9\linewidth]{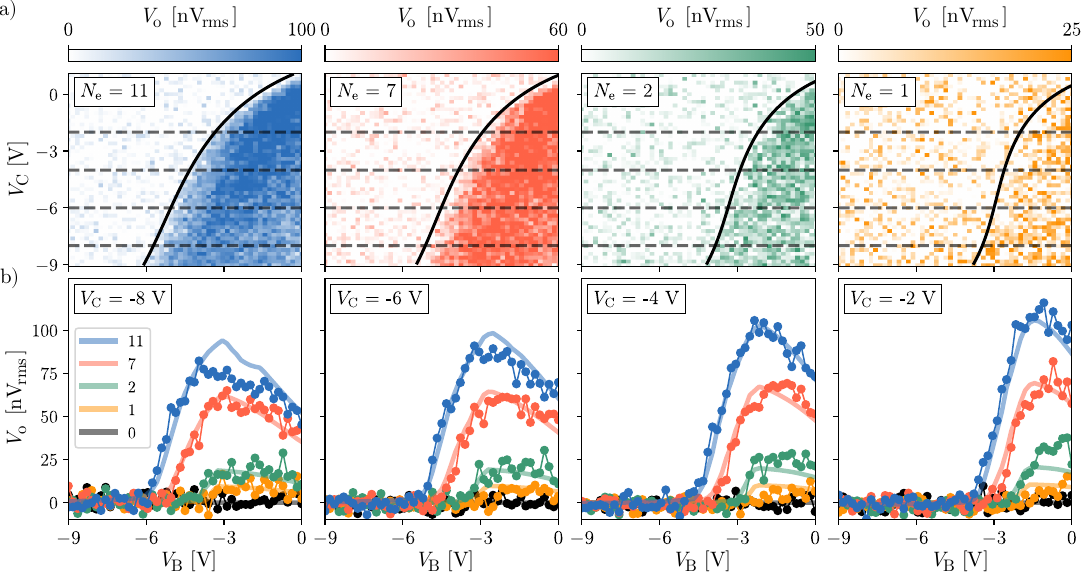}
	\caption{Controlling the number of electrons per packet. (a) Measured $V_\mathrm{O}$ for varying Barrier ($V_\mathrm{B}$) and Confining ($V_\mathrm{C}$) voltages, following the same method as in Fig.~\ref{fig:SensorExplanation}. Panels left to right show measured data for decreasing electron packet size. The solid lines are simulated current onset boundaries for $N_\mathrm{e}=11, 7, 2$ and $1$ as indicated in the boxes. (b) Cut-throughs of the data shown in (a) taken at $V_\mathrm{C}$ = -8, -6, -4 and -2 V (left to right). These voltage values are indicated by the dashed lines in (a). Also shown in black are data recorded for the case in which all electrons are ejected from the packet sensor. The solid lines are simulation results for $N_\mathrm{e}=11, 7, 2, 1$ and 0.} 
\label{fig:SingeElectronData}
\end{figure*}

We now explore the limits on how few electrons can be reliably confined within each charge packet. We achieve this by repeatedly ejecting a fraction of the electrons from the charge packets. As $N_\mathrm{e}$ is reduced, the magnitude of the ac transport signal decreases and the current onset boundary shifts. Comparison with simulations allows us to infer the number of electrons remaining in the packets.

The process of reducing the number of electrons in the packets begins by unloading the packets from the sensors back into the CCD array. This is accomplished by setting the packet sensor voltages negative, from left to right, while biasing the nearest CCD gate to the right of the sensor positive. Once out of the sensor, the packet is clocked back to the intersection seen in Fig.~\ref{fig:PacketInit}. By adjusting the voltages on both the CCD array and the Reservoir Door, some electrons can be spilled from the packets back into the Reservoir. This packet reduction procedure is similar to that described in our previous work\cite{castoria2025sensing}. For the data shown in Fig.~\ref{fig:SingeElectronData}, the Reservoir was biased to +2 V, the upper metal layer was biased to -6 V, the gates containing the packets and the Reservoir Doors were biased to -4 V, and the other two gate phases were biased to progressively more negative values for each iteration. This value was -5 V, -6 V, -7 V, -7.5 V, and -8.5 V, for the data corresponding to $N_\mathrm{e} = 11, 7, 2, 1, 0$ respectively.

In Fig.~\ref{fig:SingeElectronData}(a) we show the ac transport signal $V_{\mathrm{O}}$ against the Barrier bias voltage and the Drive and Sense dc voltage, for four different electron packet sizes (panels left to right). In each plot we also show the current onset boundary determined by simulation. For the four plots (left to right) we find the values $N_\mathrm{e}=11, 7, 2$ and $1$ give current boundaries which best match the experimental data. As expected, the current onset moves to less negative (positive) values of $V_\mathrm{B}$ ($V_\mathrm{C}$) as the electron number decreases. We were unable to observe any further shift of the boundary beyond that shown in the right-most panel; instead, when attempting to reduce the electron number further the signal became too small to measure. This behavior is as expected for the limit when $N_\mathrm{e} = 1$.

In Fig.~\ref{fig:SingeElectronData}(b) we show an alternative analysis using cut-throughs of the data presented in (a). Here the panels (left to right) show $V_{\mathrm{O}}$ as a function of Barrier voltage for $V_{\mathrm{C}} = -7, -5, -3$ and $-1$ V. Data for the four separate electron packets and for the case in which all electrons were removed from the packet sensor ($N_\mathrm{e}=0$) are shown in each panel. We also plot simulation results for the same conditions, again using the values $N_\mathrm{e}=11, 7, 2, 1$ and $0$. 

The consistent agreement between experiment and simulation shown in Fig.~\ref{fig:SingeElectronData} indicates that our determination of the electron packet size is accurate. We therefore conclude that for the case shown in the right-most panel of Fig.~\ref{fig:SingeElectronData}(a) each packet sensor holds a single electron, on average. However, we were unable to observe clear steps in the signal magnitude or current boundary that would correspond to discrete and simultaneous changes in the number of electrons in each sensor. This is likely a consequence of our measurement scheme in which the charge measurement is performed for 34 channels in parallel. In future devices, unambiguous determination of the electron number \textit{in a single sensor} can be aided by increasing the electron capacitive coupling to the underlying Sense gate (for example by reducing the helium film thickness) and by reducing the stray capacitance of the measurement line. We estimate that gains in signal magnitude of $\times10$$-$$100$ should be readily achievable, allowing HEMT amplifier-based single eHe detection without the parallel measurement scheme employed here.      

\section{Selective Electron Shuttling}
\label{sec:clocking}

\begin{figure*}
\includegraphics[width=0.9\linewidth]{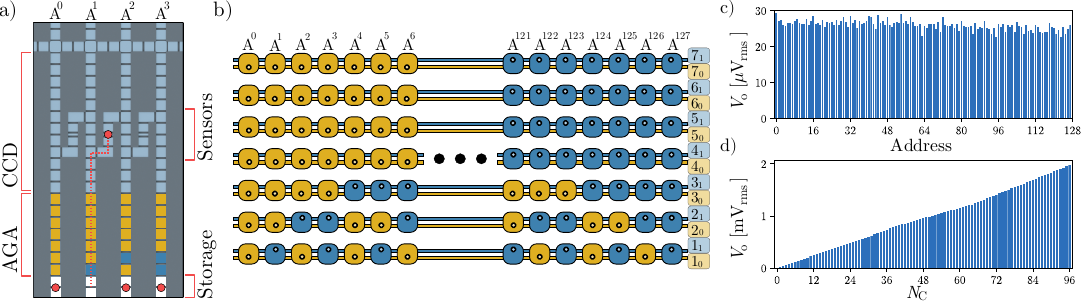}
\caption{Selective electron shuttling. (a) Render of a unit cell. At the bottom of the image are the two storage electrodes (white) filled with electron packets (red circles). Above them are the seven AGA gates (orange or dark blue), followed by CCD gates and three packet sensors (light blue). In this image, the packet in address column $A^1$ has been clocked along the red dotted path into a sensor for measurement. (b) AGA gate wiring schematic. The labels for the encoded addresses are above and the labels for the 14 control lines are on the right. Vias connecting each gate to one of the control lines allow each column to encode one of the $2^{N}$ possible addresses, where $N$ is the number of gates in a column. (c) Sensor voltage measured for sequential electron packet shuttling from each of the 128 AGA. Each address is loaded from the storage electrodes to the sensor as depicted in (a), then returned to the storage electrodes after measurement. (d) Sensor voltage measured while cumulatively loading all of the sensors. 96 of the 128 addresses are iterated over to load all of the shuttling zone packet sensors. However in this case, the packets are not returned to their storage locations after loading.}
\label{fig:Addressing}
\end{figure*}

Having characterized the sensitivity and voltage bias dependence of the Wonder Lake packet sensors, we now implement them to demonstrate selective electron shuttling. The shuttling measurements are performed in the Shuttling Zone adjacent to the electron Reservoir. As described in Section~\ref{sec:Device}, the Shuttling Zone features a unit cell which contains an array of microchannels and control electrodes. The unit cell is shown schematically in Fig.~\ref{fig:Addressing}(a). Each cell contains four vertical microchannels, the bottom of which are patterned with CCD gates for electron shuttling, which are linked horizontally by three zig-zag shaped microchannels. At the center of these zig-zag microchannels are electron packet sensors identical to those described in Section~\ref{sec:sensor}.

The unit cell is repeated 32 times in the horizontal direction across the device, giving a total of 128 vertical channels and 96 packet sensors. As explained below, each of the 128 vertical gate arrays is wired uniquely, allowing electron shuttling to be selectively performed in any of the 128 microchannels. We are therefore are able to load electrons from a single vertical channel into only one of the 96 packet sensors for charge readout. However, to boost our signal-to-noise ratio in a manner similar to the scheme employed in Section~\ref{sec:sensor}, the horizontal set of unit cells is copied 34 times in the vertical direction, with all CCD gates and sensor electrodes wired in parallel across the 34 copies. 


An efficient arrangement of control electrodes allows us to use only 14 control voltages to select from which of the 128 vertical channels an electron packet can be shuttled to the packet sensor for readout. This selective shuttling is performed using 7 of the gate electrodes in each vertical channel, which we term the \emph{Addressable Gate Array} (AGA). The wiring scheme is shown in Fig.~\ref{fig:Addressing}(b). Beneath the 7 gates of the AGA, 14 wires run horizontally with pairs positioned under each gate. Vias between the metal layers connect each gate in the AGA to one of the two underlying wires. This enables each AGA to be wired in one of $2^7=128$ unique configurations, one for each column. Using $0$ or $1$ to signify whether each gate is connected to the lower or upper of the underlying pairs, each of the 128 AGA can be described using a unique 7-digit binary number. We label each of the AGA by an address A$^\beta$ where $\beta$ is the decimal representation of this 7-digit binary number, as shown in Fig.~\ref{fig:Addressing}(b). For example, A$^{42}$ denotes the column with wiring described by the binary number 0101010, signifying that for this column the AGA gates are wired alternately to the lower and upper of the control voltage pairs.

The selective electron shuttling proceeds as follows. Electron packets are loaded from the Reservoir, along the horizontal CCD gate arrays, into each of the 128 microchannels. The packets are initially localized above two gates which form \textit{storage} zones, as shown at the bottom of each channel in Fig.~\ref{fig:Addressing}(a). All 14 AGA control lines are initially biased to -5 V. The 3-phase voltage sequence is then applied to the unique set of seven control lines associated with a selected AGA. For the electrons in other columns, at least one gate of the AGA is therefore not included in the pulse sequence and remains at -5 V. The resulting potential barrier blocks electron transport, and the electron packet remains trapped at the gate behind the barrier. Once the selected electron packet is fully clocked out of its AGA, all of the control lines are reset to -5V, from the top gates down. This clocks the electrons in all other columns back to their respective storage zones. 

 After this shuttling sequence is completed, the electron packet from the selected column is moved into the main CCD array and then into the closest packet sensor for measurement. For address $A^1$, the path from storage zone to sensor is shown as the red dashed line in Fig.~\ref{fig:Addressing}(a). In Fig.~\ref{fig:Addressing}(c) we show the change in the packet sensor signal $V_\mathrm{O}$ as electron packets are clocked in turn from the 128 AGAs into the sensors. Here the Confinement gates and Barrier gates are biased at -6~V and 0~V respectively. A signal is detected for each address number, confirming that electron packets are successfully loaded from each channel. Although these measurements are conducted in parallel over the 34 vertical replicas of the AGA system, the magnitude of the signal is very similar for all channels, on average about $26 \mu\mathrm{V}$, which corresponds to individual packets of 45 electrons. The conversion from voltage to electron packet size was calculated in the same way as in Section~\ref{sec:single}, but using the relevant quantities for the second HEMT amplifier circuit. 

 With 96 sensors per row, and 34 vertical replicas, there are a total of 3264 individual packet sensors in the electron shuttling region, which are all wired in parallel. Therefore, the only distinction between the signals shown in Fig.~\ref{fig:Addressing}(c) is the clock sequence that determines from which address packets are shuttled. Therefore, it is important to confirm that the 128 signals arise from distinct packets stored in different addresses. Fig.~\ref{fig:Addressing}(d) shows the sensor signal as the sensors are cumulatively loaded. Here we iterate over 96 of the 128 addresses to load packets into all of the 96 sensors per row. However, after each packet is loaded, it is not returned to its storage zone. Now, the signal grows as the number of sensor columns loaded, $N_\mathrm{C}$, increases. Together, the measurement routines shown in Figs.~\ref{fig:Addressing}(c) and (d) demonstrate the ability to control and address distinct electron packets throughout the shuttling zone.

Once selectively transported through the AGA from its initial storage zone, an electron packet can be deterministically moved around the 2D Wonder Lake geometry. For example, after sensing is performed, the packet can be moved out of the packet sensors and sent to any of the neighboring 127 microchannels or 95 packet sensors. Also, an additional electron packet can be selected and moved through the AGA, and independently transported to meet the first electron packet, for example at one of the packet sensors. The two packets can then be merged and after some time split again, using voltages applied to the Drive, Sense and Barrier electrodes. These operations mimic the functionality required in future quantum processors, where spin qubits must be selectively transported to readout zones or brought together to perform quantum gate operations\cite{lyon2006spin}. In this way, the Wonder Lake device serves as a prototype quantum processor in which deterministic electron transport operations can be demonstrated.  

\section{Conclusions}
\label{sec:conclusions}

In summary, we have demonstrated selective two-dimensional electron shuttling across a superfluid helium film covering the Wonder Lake CMOS control chip. This enables deterministic transport of discrete electron packets through a two-dimensional network of microchannels. The device architecture supports 128 independently addressable transport channels, each capable of performing CCD-style electron transport between storage zones and electron packet sensors. The electron shuttling paths are many tens or hundreds of micrometers in length, and shuttling sequences can be repeated at frequencies exceeding $\sim100$ kHz with no observable electron loss during the typical measurement timescale of many hours. In addition, by systematically reducing the number of electrons in each packet we show that control at the level of single electrons can be achieved. This level of performance, achieved through optimization of our CMOS electrode design and the intrinsic cleanliness of the helium substrate, establishes a robust platform for large-scale, high-fidelity electron manipulation. 

This on-demand electron shuttling demonstrates the suitability of the eHe system for future quantum processors in which coherent electron transport allows all-to-all qubit connectivity within a scalable 2D architecture. This qubit reconfigurability is ideally suited to the implementation of surface codes, which require local two-qubit operations between neighboring data and ancilla qubits in a planar array, and offer the high fault-tolerance thresholds necessary for a practical quantum computing system\cite{Fowler2012ErrorCorrection, dennis2002topological}. The combination of expected long electron spin coherence times, geometric compatibility with 2D error-correcting lattices, and flexible qubit transport makes eHe a promising platform for realizing large-scale quantum processors. Therefore, the precise electron control performed using our Wonder Lake device lays a strong foundation for future eHe quantum computing technologies.

\begin{acknowledgments}
We thank G. Koolstra and M. M. Feldman for useful discussions and J. Theis for technical support. This work made use of the Pritzker Nanofabrication Facility of the Institute for Molecular Engineering at the University of Chicago, which receives support from Soft and Hybrid Nanotechnology Experimental (SHyNE) Resource
(NSF ECCS-2025633), a node of the National Science
Foundation’s National Nanotechnology Coordinated
Infrastructure.
\end{acknowledgments}

\bibliography{Electron_Shuttling} 







\end{document}